\documentstyle[aps,prl,twocolumn,epsfig,float]{revtex}
\begin{document}
\draft

\twocolumn[\hsize\textwidth\columnwidth\hsize\csname@twocolumnfalse\endcsname

\title{A Nonperturbative Approach to One-Particle Green's Function}
\author{Jongbae Hong}
\address{Department of Physics Education,\\
         Seoul National  University, Seoul 151-742, Korea}

\maketitle
\begin{abstract}
A nonperturbative method to obtain on- and off-site one-particle Green's 
function is introduced and applied to noninteracting Hubbard model with 
next nearest neighbor hopping and interacting Hubbard model in large dimensions, 
for example. The former gives some lessons on the method and 
shows the advantage of the method compared with continued fraction formalism. 
The latter is treated by selecting important dynamic processes contributing 
to the Green's function when correlation is strong. We consider the model 
in the Bethe lattice with large connectivity. The dynamic processes 
describing on-site spin fluctuation clearly shows metal-insulator transition 
in the paramagnetic ground state at half filling.  

\end{abstract}

\pacs{PACS numbers : 11.15.Tk,71.10.Fd,71.30.+h}
]

\pagebreak

One of recent interesting subjects in condensed matter physics is the 
phenomena of the strongly correlated systems. 
A typical characteristic  of the strongly  correlated system is the hardness 
of getting reasonable 
solution in  terms of the traditional  perturbative method.  
The  Hubbard model, for instance, is one of typical strongly  correlated 
systems and is recently revisited with great interest as a possible model 
for high temperature superconductor.\cite{anderson,dagotto} 
This model has not been  solved exactly except some of 
statics in one dimension via Bethe ansatz  even though it looks simple. 
Recently, however, the statics\cite{muller} and dynamics 
\cite{jarrell0,gk,georges0,rozenberg0,pruschke,jarrell} of the Hubbard model 
have been treated exactly in large dimensions, where mean-field approximation 
is valid, with help of numerical techniques.
These exact  solutions show interesting behaviors such as band collapsing and 
quasiparticle mass enhancement near 
Mott-Hubbard transition, which usual perturbative  method cannot show. 
In finite dimensions, however, only numerical methods such as exact diagonalization 
and quantum Monte Carlo method have shown 
some meaningful results\cite{dagotto} even though they have severe 
limitations. Therefore, an effective nonperturbative method is badly required. 

This Letter reports a novel method to calculate  one-particle
Green's function nonperturbatively. The Green's function is usually 
obtained by summing Feynman diagrams\cite{fetter} or 
truncating equation of motion for the Green's function.  The 
former is well-known and solid in its foundation, but it is perturbative 
in principle and the latter suggested by 
Zubarev\cite{zubarev} needs truncation at very low level. The famous works   
of Hubbard\cite{hubbard} have been done using the latter method. 
But the results are only partly acceptable. The method we present here is 
very simple in its form and has never been used before as  far as we know.  
We test the effectiveness of the method by solving finite-size 
noninteracting Hubbard model and interacting Hubbard model in large dimensions. 

In this work, we treat the retarded Green's function which is related to other 
types of Green's functions such as time-ordered and thermal Green's function 
and to physical quantities such as single-particle density of states (DOS) and 
internal energy directly. 
The well-known one-particle retarded Green's function in real space is 
\begin{equation}
G_{jk}^{\rm R}(t)=-i\theta(t)
\langle\{c_{j\sigma}(t),c_{k\sigma}^{\dag}\}\rangle
\label{gr}
\end{equation}
and its Fourier transform is
\begin{equation}
  G_{jk}^{\rm R}(\omega +i\epsilon )=-\frac{i}{2\pi}\int_{0}^{\infty}
    \langle \{c_{j\sigma}(t), c_{k\sigma}^{\dag}\}\rangle
    \exp^{i\omega t-\epsilon t}dt,
    \label{green}
\end{equation}
where the superscript R denotes the notation of the retarded Green's
function. Let's notice that the main part of Eq. (1) is the projection 
of the time evolution $c_{j\sigma}(t)$ onto $c_{k\sigma}$ in the Liouville 
space whose inner product is defined as 
\begin{equation}
(f,g)\equiv\langle\{f,g^{\dag}\}\rangle ,
\label{inner}
\end{equation}
where $f$ and $g$ are vectors of the Liouville space and 
the angular brackets mean ensemble average, and Eq. (2) is the Laplace 
transform of $\langle\{c_{j\sigma}(t),c_{k\sigma}^{\dag}\}\rangle$. 
Thus our interest to get the Green's function (1) or (2) is in
the projection $\langle \{c_{j\sigma}(t), c_{k\sigma}^{\dag}\}\rangle$.\cite{fulde}

Let us assume that we have a complete set of orthogonal bases 
$\{e_{\nu}|\nu =1,2,\cdots ,\infty\}$ where $e_1=c_{j\sigma}$, which spans 
the Liouville space describing the dynamics of $c_{j\sigma}(t)$. Thus the 
time evolution $e_1(t)=c_{j\sigma}(t)$ can be described in terms of 
these bases,
\begin{equation}
 e_1 (t)= \sum_{\nu=1}^{\infty}A_\nu (t) e_\nu.
\label{feq}
\end{equation} 
Here $A_1(t)=(e_1(t),e_1)/(e_1,e_1)=\langle\{c_{j\sigma}(t), 
c_{j\sigma}^{\dag}\}\rangle$ is just the projection we are interested 
in Eqs. (1) and (2). To get $A_1(t)$, we apply the operator identity 
$\partial /\partial t=iL$ to Eq. (\ref{feq}), 
where $L$ is the Liouville operator and we set $\hbar=1$. Then one obtains
\begin{equation}
\sum_{\nu=1}^{\infty} \dot{A}_{\nu}(t)e_\nu 
=\sum_{\nu=1}^{\infty} A_\nu (t)iLe_\nu .
\label{aeq}
\end{equation}                       
Taking inner product in both sides of Eq. (\ref{aeq}) 
with $e_{\mu}$, one gets a system of linear equations for  $A_\nu (t)$ 
such that
\begin{equation}
\dot{A}_{\mu}(t)= \sum_{\nu=1}^{\infty}M_{\mu\nu}A_\nu (t)
\label{anu}
\end{equation}
for $\mu=1,2,\cdots ,\infty$, where $M_{\mu\nu}= 
\frac{(iLe_\nu,e_\mu)}{(e_\mu,e_\mu)}$. 
Laplace transform of Eq. (\ref{anu}) is written as 
$zA_1(z)-1=\sum_{\nu=1}^{\infty}M_{1\nu}A_{\nu}(z)$ for $\mu=1$
and $zA_\mu(z)=\sum_{\nu=1}^{\infty}M_{\mu\nu}A_{\nu}(z)$ 
for $\mu\geq 2$. Use of the 
boundary conditions $A_1 (t=0)=1$ and $A_\nu (t=0)=0$ for $\nu \ge 2$ 
has been made. These equations can be represented in a matrix form 
\begin{equation}
(z{\rm {\bf I}-{\bf M}}){\rm {\bf A}_{c}={\bf B}_{c}},
\label{matrix}
\end{equation}
where ${\bf I}$ is unit matrix and ${\rm {\bf A}_{c}}$ and ${\rm {\bf B}_{c}}$ 
are column matrices with elements 
$A_1(z),A_{2}(z),A_{3}(z),\cdots$ and $1,0,0,\cdots$, respectively. 
Using Cramer's rule, one gets  
\begin{eqnarray}
A_{1}(z)&=&\frac{{\mbox {cofactor of}} \; (z{\rm {\bf I}-{\bf M}})_
{11}}{det(z{\rm {\bf I}-{\bf M}})}   \\ 
\label{det}
&=&\int_0^{\infty}\langle \{c_{j\sigma}(t),c_{j\sigma}^{\dag}\}\rangle
\exp^{-zt}dt \nonumber
\end{eqnarray}  
and 
\begin{eqnarray}
A_\nu(z)&=&\frac{{\mbox {cofactor of}} \; (z{\rm {\bf I}-{\bf M}})
_{1\nu}}{det(z{\rm {\bf I}-{\bf M}})} \\
\label{off}
&=&\frac{1}{\langle\{e_{\nu},e_\nu^{\dagger}\}\rangle}\int_0^{\infty}
\langle \{c_{j\sigma}(t),e_{\nu}^{\dag}\}\rangle\exp^{-zt}dt \nonumber
\end{eqnarray} 
for $\nu\geq 2$, where numerators are cofactors of the elements 
in $det(z{\rm {\bf I}-{\bf M}})$. 

It  is  well-known  that  the  single-particle  DOS   
$\rho_\sigma(\omega)$  is  given  by the
one-particle retarded Green's function at the same site\cite{hubbard}, i.e.
\begin{eqnarray}
\rho_\sigma (\omega)&=&-\frac{2}{N}\lim_{\epsilon \rightarrow 0^+}
\sum_{j} {\rm Im} G_{jj}^{\rm R}(\omega +i\epsilon) \\
\label{single}
&=&\frac{1}{\pi}\lim_{\epsilon\rightarrow 0^+}
{\rm Re}A_1(z)|_{z=-i\omega+\epsilon}. 
\label{dos}
\end{eqnarray} 
On the other hand, the off-site retarded Green's function 
$G_{jk}^{\rm R}(\omega +i\epsilon)$ is given by $A_\nu(z)$ of Eq. (9) when 
$e_\nu=c_{k\sigma}$. Therefore, our task to get one-particle retarded 
Green's function is to construct orthogonal bases describing the dynamics 
of the fermion operator, $c_{j\sigma}(t)$, first and 
calculate the matrix elements $M_{\mu\nu}$ of Eq. (\ref{matrix}). 

 This is a right place to give a remark on this formalism. One may consider 
the operators $(iL)^\nu e_1$ as a complete set of linearly independent 
bases and construct orthogonal bases in terms of Gram-Schmidt process. 
Then we have the following recurrence relation\cite{mori,lee,hong0}
\begin{equation}
e_{\nu +1}= iLe_{\nu}-\alpha_\nu e_\nu+\Delta_\nu e_{\nu-1},
\label{rr}
\end{equation}
where $e_{0}\equiv 0, \alpha_\nu=\frac{(iLe_\nu,e_\nu)}{(e_\nu,e_\nu)}$ 
and $\Delta_{\nu}=\frac{(e_{\nu+1},e_{\nu+1})}{(e_{\nu},e_{\nu})}$,
when the inner product satisfies the relation $(iLe_\nu ,
e_\mu)=-(e_\nu ,iLe_\mu)$ which implies the time-reversal symmetry of the 
inner product. Therefore, when inner product contains equilibrium average
such as Eq. (\ref{inner}), this relation is always satisfied.

The recurrence relation (\ref{rr}) makes the matrix elements $M_{\mu\nu}$
vanish for $|\nu-\mu|\ge 2$, since inner products $(iLe_\nu,e_\mu)$ vanish, 
and this reduces the determinants of Eq. (8) to tridiagonal forms. 
Then $A_1(z)$ can be written as a well-known infinite continued 
fraction\cite{fulde,mori,lee,hong0}
\begin{equation}
A_{1}(z)=\frac{1}{z-\alpha_1+\frac{\Delta_1}
{z-\alpha_2+\frac{\Delta_2}{\ddots}}}.              
\label{conti}
\end{equation}
However, it is practically hard to construct orthogonal bases using 
$(iL)^\nu e_1$ and impossible to get off-site Green's function. 
More recommendable way is to find linearly independent bases intuitively with
help of $(iL)^\nu e_1$ which generates all kinds of linearly independent 
operators. 

The author and  Kee\cite{hong1} have studied  the Hubbard model  using recurrence relation 
(\ref{rr}) and continued   fraction (\ref{conti})   in terms  of  the  bases  obtained  by  
large-$U/t_*$ expansion. 
Even though the bases were orthogonal, they did not satisfy the recurrence relation for whole 
region of $U/t_*$. Therefore, the continued fraction formalism cannot describe the  physics of 
intermediate 
and   low   $U/t_*$    regime.   For   this   reason,    the   experiment   for   insulating 
$V_2O_3$\cite{thomas} 
has been explained very well.\cite{hong2} Similar method has been applied to doped 
Hubbard model with large-$U/t_*$  and shown characteristic features  of La- and Nd-based 
doped cuprates quite well.\cite{hong3} To describe the transition regime well one must use present 
formalism instead of continued fraction (\ref{conti}).

We present here a tutorial example of our formalism using
noninteracting Hubbard model on 4-site lattice with next nearest neighbor hopping, i.e.
\begin{eqnarray}
H_0&=&-t(c_1^{\dagger}c_2+c_2^{\dagger}c_3
+c_3^{\dagger}c_4+c_4^{\dagger}c_1+{\it h.c.}) \nonumber \\
&+&t^{\prime}(c_1^{\dagger}c_3+c_2^{\dagger}c_4+{\it h.c.}).
\label{4site}
\end{eqnarray}
We first consider $t^{\prime}=0$ case.                                                   
One naturally selects $c_1, c_2, c_3, c_4$ as 4-orthogonal bases $e_1, e_2, e_3, e_4$. 
Using these bases Eq.(8) gives $A_1(z)=(z^2+2t^2)/[z(z^2+4t^2)]$ 
and $\rho(\omega)=\frac{1}{2}\delta (\omega)+\frac{1}{4}\delta 
(\omega -2t)+\frac{1}{4}\delta (\omega +2t)$. On the other hand,
$A_2(z)$ and $A_3(z)$ give off-site Green's functions such as 
$G_{12}^{\rm R}(z)=\frac{t}{2\pi}\frac{1}{z^2+4t^2}$ 
and $G_{13}^{\rm R}(z)=\frac{it^2}{\pi(z^3+4t^2z)}$, respectively.
One can also easily construct orthogonal bases using $(iL)^\nu c_1$, i.e., 
Eq.(\ref{rr}). If we choose $e_1=c_1$,
then Eq. (\ref{rr}) gives $e_2=it(c_2+c_4)$, 
$e_3=-2t^2c_3$, and $e_4=0$.  $(iL)^\nu c_1$ 
for $\nu\geq 3$ are no longer linear independent vectors.
Eq.(\ref{conti}) and Eq.(\ref{dos}) give rise to the same $A_1(z)$ 
and $\rho(\omega)$ as above. However, this does not give off-site Green's function.

Including next nearest hopping discriminates two methods. The first approach raises no 
difficulty in treating next nearest hopping, while the second raises big complexity in
constructing orthogonal bases. The first method easily gives $A_1(z)=[z^3+(2t^2+t'^2)z
+2it^2t']/[z^4+(4t^2+2t'^2)z^2+8it^2t'z-4t^2t'^2+t'^4]$ from Eq. (8).

This trivial example gives us some lessons about this method. First,   
the bases represent possible independent ways of annihilating fermion at a 
particular site. Second, the bases obtained intuitively also give the same 
result as one obtained by using recurrence relation and continued fraction, 
while this does not give off-site Green's function and meets great complexity when 
next nearest neighbor hopping is included. Third, one can figure out the 
meaning of bases, e.g. $t^2c_3$ means annihilation at site 1 after twice hopping 
from site 3. Therefore, one can construct bases describing $c_j(t)$ by considering 
the roles of bases intuitively. It is safe to be guided by Eq. (\ref{rr}) when 
one constructs independent bases intuitively.

As a meaningful example of constructing bases by considering their roles, we treat 
the half-filled Hubbard model in large dimensions. In describing $c_{j\sigma}(t)$, where
$\sigma$ denotes up-spin, one must consider all possible processes while an up-spin 
electron at site $k$, which represents all lattice sites including $j$, hops, arrives 
at site $j$, and annihilates. For the strongly correlated case ($U>t$), where $U$ and $t$ 
are on-site Coulomb interaction and hopping integral, respectively, the down-spin 
number fluctuation (on-site spin fluctuation) when up-spin exists at the same site is the 
most effective part in a basis representing the process of the same order of energy, 
because it yields $U$ for each fluctuation.

We consider the Bethe lattice with arbitrarily large connectivity, which is rather easy since
the mean-field   approximation is   valid.  We  show   the metal-insulator   transition in    
the 
paramagnetic 
ground state for the half-filled Hubbard model by calculating the single-particle 
density of states (\ref{dos}).  
This is the well-known phenomenon studied by Hubbard\cite{hubbard} and Brinkman and 
Rice\cite{brinkman} long time ago. Recently the infinite dimensional model has been 
studied exactly by several authors.\cite{georges0,rozenberg0,pruschke,jarrell}

The interacting Hubbard model is written as 
\begin{equation}
   H=-\sum_{j,k,\sigma}t_{jk}c_{j,\sigma}^{\dag}c_{k,\sigma}+
   \frac{U}{2}\sum_{j,\sigma}n_{j,\sigma}n_{j,-\sigma}.
\label{hub}
\end{equation}                               
We consider only nearest neighbor hopping. To construct bases, we choose $e_1=c_{j\sigma}$ 
as before and consider down-spin fluctuation at each site where up-spin electron hops. 
We collect the dominant processes describing down-spin number fluctuation 
and neglect   hopping dominant   processes such   as hopping  without  down-spin  number 
fluctuation.
The bases are written as
\begin{eqnarray}
 &e_{2\nu-1}&=(-U)^{\nu-1}\overbrace{\Sigma_{k}'\cdots\Sigma_{p}'}^{\nu-1} 
     \overbrace{t_{jk}\cdots t_{pq}}^{\nu-1} \nonumber \\ 
     &\times&\!\!\!
     \overbrace{(\delta n_{j,-\sigma}+\delta n_{q,-\sigma})\cdots
     (\delta n_{p,-\sigma}+\delta n_{q,-\sigma})}^{\nu-1} c_{q,\sigma},
\label{vec1}           \\
 &e_{2\nu}&=i(-U)^{\nu}\overbrace{\Sigma_{k}'\cdots \Sigma_{p}'}^{\nu-1}
  \overbrace{t_{jk}\cdots t_{pq}}^{\nu-1} \nonumber \\ 
     &\times&\!\!\!
    \overbrace{ (\delta n_{j,-\sigma}+\delta n_{q,-\sigma})\cdots
     (\delta n_{p.-\sigma}+\delta n_{q,-\sigma})}^{\nu-1}
     \delta n_{q,-\sigma}c_{q,\sigma},
\label{vec2}
\end{eqnarray}
for $\nu \ge 1$, where $\delta n_{j,-\sigma}=n_{j,-\sigma}-\langle n_{j,-\sigma}\rangle$, 
$\nu$ above the overbrace denotes  the number of the same symbols, 
and the prime means that all site indices under summation sign are different each other.
The  orthogonality  of   above bases   is  retained   by site-difference   and  $\langle\delta 
n_{i,-\sigma}
\rangle=0$. The property $\delta n_{i,-\sigma}^2=\langle\delta n_{i,-\sigma}
\rangle^2=1/4$ for half-filling is used.

The bases of Eqs. (\ref{vec1}) and (\ref{vec2}) do not satisfy the recurrence relation (\ref{rr}) 
even though they are orthogonal each other. Therefore, one cannot use the continued fraction 
formalism (\ref{conti}). One must use more unrestricted formalism (8) to calculate one-particle
on-site Green's function.  The matrix  elements  $M_{\mu\nu}$ are  obtained as  follows in 
terms of the definition of inner product (\ref{inner}) and the mean-field 
approximation neglecting spatial correlations, e.g.
$\langle n_{j,\sigma}n_{k,\sigma}\rangle=\langle n_{j,\sigma}\rangle\langle n_{k,\sigma}\rangle$ 
for $j \neq k$, which is valid in the limit of large connectivity: 
\begin{eqnarray}
 &M_{\nu,\nu}&=-\frac{iU}{2}, \hspace{0.5cm} M_{2\nu-1,2\nu}=-\frac{U^2}{4}, \hspace{0.5cm}  
M_{2\nu,2\nu-1}=1, \nonumber \\ 
 &M_{2\nu,2\nu+1}&=-\frac{t_{*}^2}{2}, \hspace{1cm} M_{2\nu+1,2\nu}=\frac{1}{2},  \nonumber 
\\ &M_{2\nu-1,2\nu+2}&=\frac{U^2 t_{*}^2}{8}, \hspace{0.5cm} M_{2\nu+2,2\nu-1}=-\frac{2}{U^2} 
\end{eqnarray}
for $\nu\geq 1$, where we scale $t$ as $t_*/\sqrt{2q}$, where $q$ is the coordination
number of Bethe lattice, to make the kinetic energy finite for large coordination. 
All other matrix elements vanish because electron leaving $j$ site cannot come back to
the original site via nearest neighbor hopping.  
In calculating the matrix elements, use of the approximation $q-1\approx q$
in addition to mean-field approximation has been made, which is good enough for 
a system with large connectivity. 
Once matrix elements $M_{\mu\nu}$ are obtained, the single-particle DOS can be given
by Eq. (\ref{dos}). 

Fig. 1 shows the results of the single-particle DOS for various $U/t_*$.
The rank of the matrix ${\bf M}$ is 101. 
The metal-insulator transition is clearly shown at $U_c=2t_*$ by showing band collapsing 
from insulating side and quasiparticle mass enhancement via disappearance of Fermi 
liquid quasiparticle band from metallic side. 

As a conclusion, we introduced a nonperturbative method calculating the one-particle  Green's 
function and applied it to a simple tutorial model and the half-filled Hubbard model on a Bethe lattice 
with large connectivity.  The   former provides    some lessons  on   this  new  method  and   
shows advantages compared with continued fraction formalism, while the latter shows interesting 
Mott-Hubbard transition   by collecting   orthogonal bases   describing important  dynamical 
processes of strong correlation. 
\begin{figure}[ht]
\epsfxsize=7cm
\centerline{\epsffile{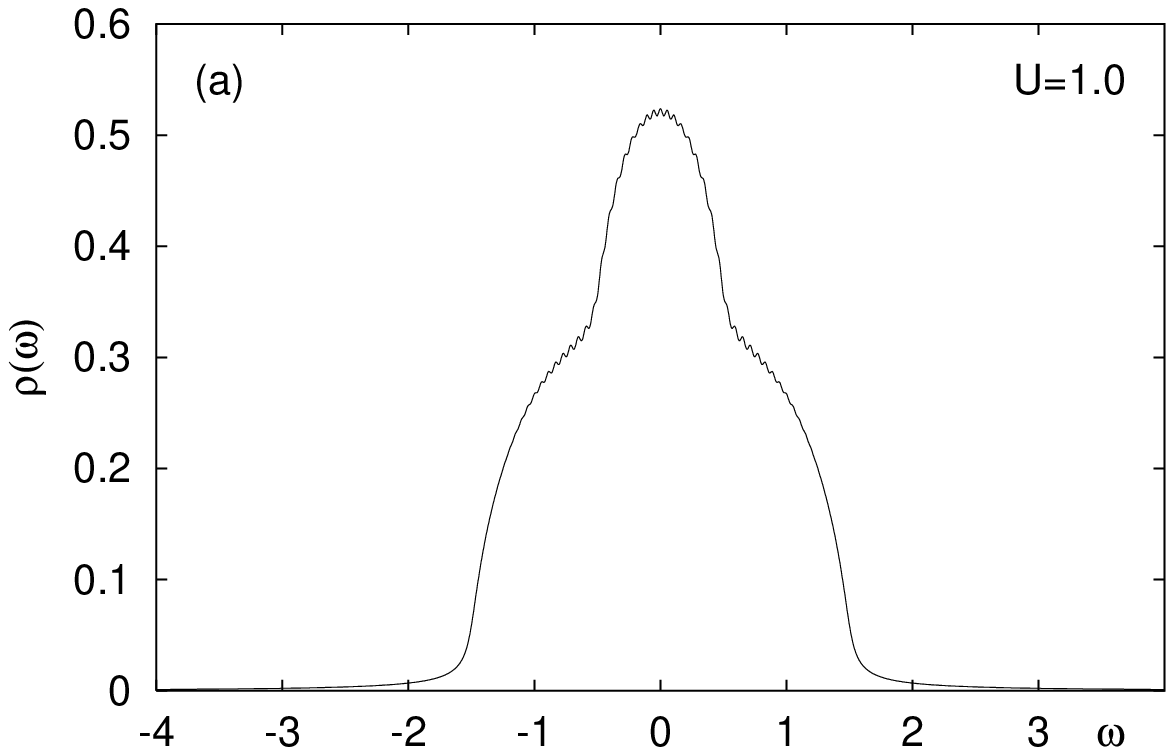}}
\centerline{\epsffile{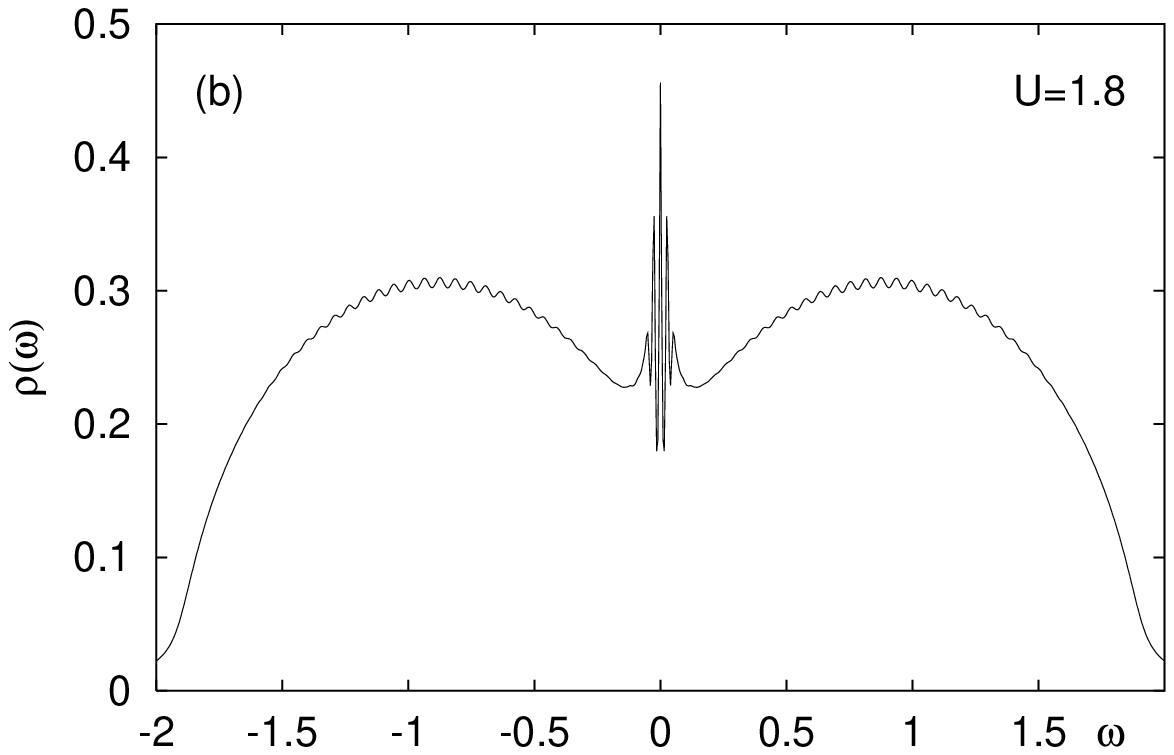}}
\centerline{\epsffile{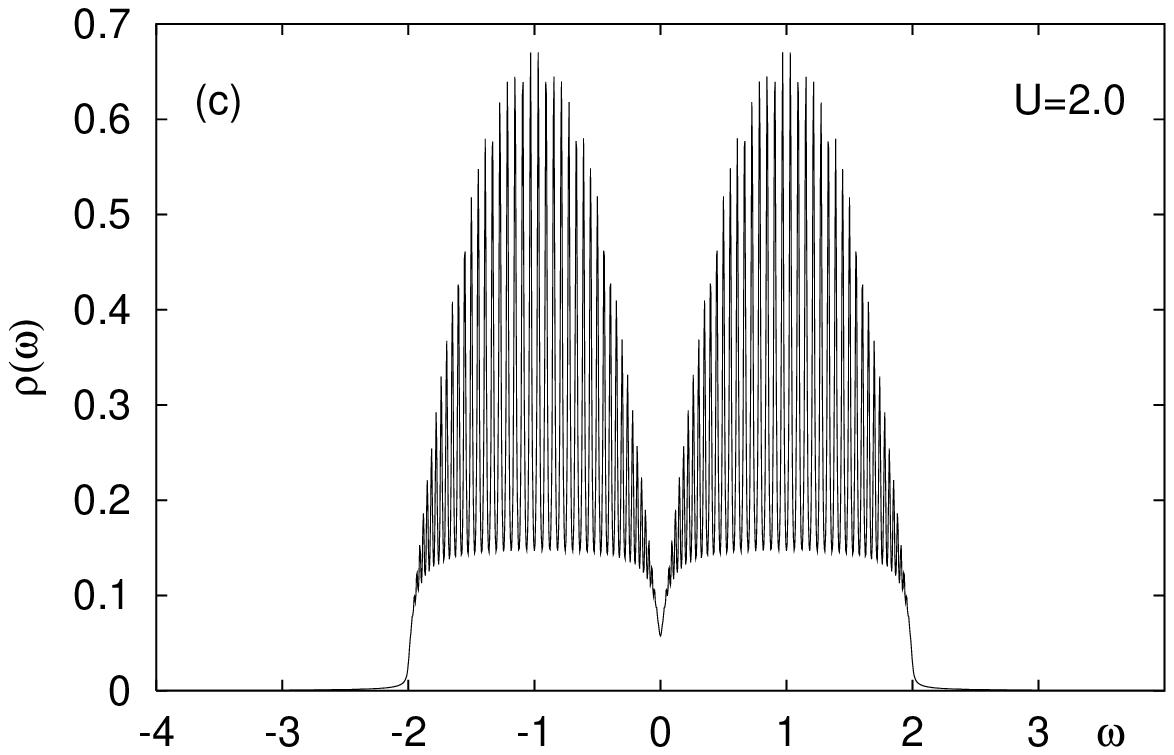}}
\centerline{\epsffile{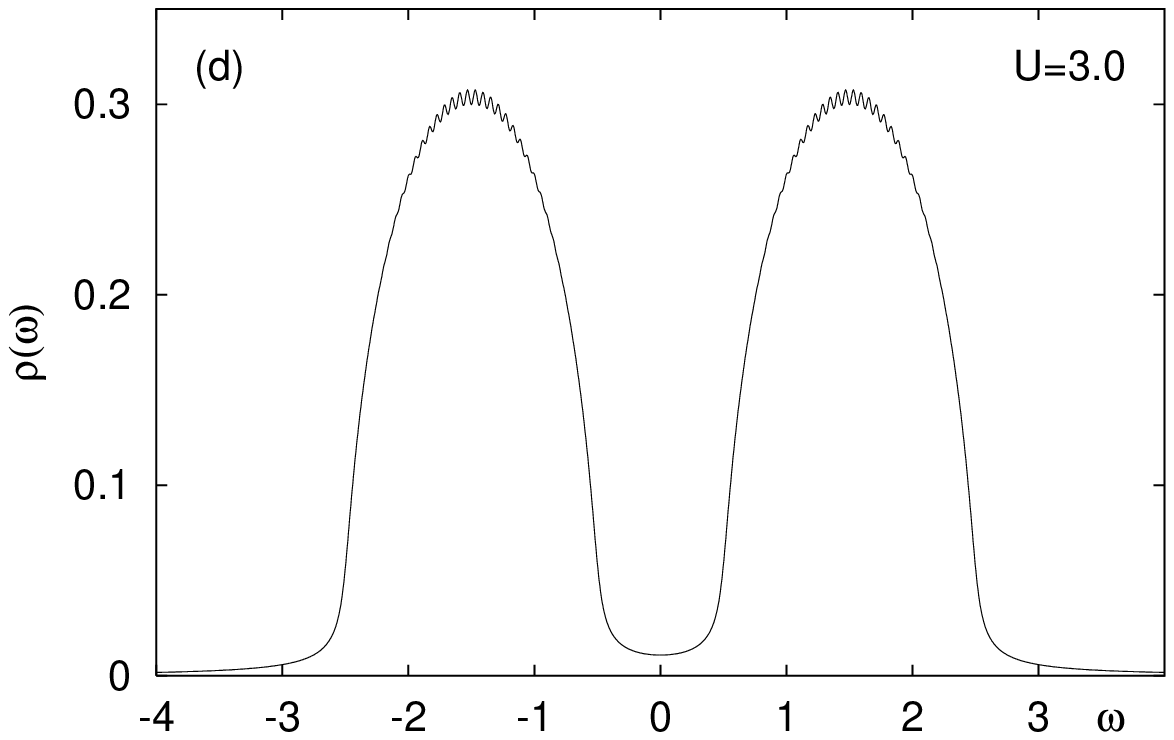}}
\caption{The single-particle DOS for various $U$. We set $t_*=1$. 
The $\delta$-function width parameter is chosen $\varepsilon=0.04$ except in the quasiparticle 
band in (b) where $\varepsilon\propto |(\omega-\epsilon_F)|^2$. (b) is magnified to show 
the band clearly.}
\label{figure1}
\end{figure}
The most crucial part of this method is the construction of orthogonal bases 
which describe the various  situation of electron  annihilation processes. These  processes are 
obviously very complicate. However,  as an  approximation, one  can select  some important  processes 
which play a major role in a particular parameter regime. The second example shows this well. Therefore,  
this formalism may  provide a possibility  to treat the  strongly correlated systems  in lower 
dimensions, which is a great challenging problem in current condensed   matter physics. 
More applications of this formalism will follow.

This work has been supported by BSRI 97-2420 of the Ministry of Education
and the KOSEF through the SRC program of SNU-CTP.

\end{document}